\documentclass[aip,jcp,reprint]{revtex4-1}

\usepackage{graphicx}
\usepackage{dcolumn}
\usepackage{bm}
\usepackage[utf8]{inputenc}
\usepackage[T1]{fontenc}
\usepackage{mathptmx}
\usepackage{etoolbox}
\usepackage{balance}
\usepackage{graphicx}

\usepackage{physics2}
\usepackage[table]{xcolor}
\usepackage{soul}
\usepackage{multirow, tabularx, booktabs}
\usepackage{braket}
\usepackage{adjustbox}
\usepackage{tablefootnote}

\usepackage{xr} 
\usepackage{threeparttable}
\usephysicsmodule{braket}

\usepackage[version=4]{mhchem}

\usepackage{physics}
\usepackage{amsmath}
\usepackage{amssymb}
\usepackage{nicefrac}
\usepackage[colorinlistoftodos]{todonotes}

\newcommand{\Tp}{{\hat{T}_{\rm pCCD}}}

\newcommand{\Lower}[1]{\smash{\lower 1.5ex \hbox{#1}}}
\makeatletter
\def\@email#1#2{%
 \endgroup
 \patchcmd{\titleblock@produce}
  {\frontmatter@RRAPformat}
  {\frontmatter@RRAPformat{\produce@RRAP{*#1\href{mailto:#2}{#2}}}\frontmatter@RRAPformat}
  {}{}
}
\makeatother

\begin{document}

\preprint{AIP/123-QED}
\newcommand*\mycommand[1]{\texttt{\emph{#1}}}


\title{Analytic Gradients and Geometry Optimization for Orbital-Optimized Pair Coupled Cluster Doubles}

\author{Saman Behjou}
\author{Iulia Emilia Brumboiu}
\author{Katharina Boguslawski}
\email{k.boguslawski@umk.pl}
\affiliation{Institute of Physics, Faculty of Physics, Astronomy, and Informatics, Nicolaus Copernicus University in Toruń, Grudziądzka 5, 87-100 Toruń, Poland}


\begin{abstract}
We introduce a reusable geometry-optimization engine in PyBEST for analytic, gradient-driven molecular structure optimization, with particular emphasis on orbital-optimized pair coupled-cluster doubles (OOpCCD/AP1roG).
The engine interfaces PyBEST with the \texttt{geomeTRIC} optimizer, combining analytic electronic-structure gradients from PyBEST with the translation--rotation--internal coordinate (TRIC) framework, step control, and convergence machinery provided by \texttt{geomeTRIC}.
Specifically, we present the first implementation of analytic OOpCCD nuclear gradients within a Lagrangian formalism.
Our approach and implementation are generally applicable to any seniority-zero wavefunctions that feature orbital optimization and allow for the evaluation of response one- and two-particle reduced density matrices.
Owing to the seniority-zero structure of pCCD and the orbital stationarity of the optimized reference, the resulting gradient equations are compact, minimizing the storage of the full two-particle reduced density matrix, and avoiding finite-difference differentiation of wavefunction parameters.
Validation on representative closed-shell systems shows that the OOpCCD-based PyBEST--\texttt{geomeTRIC} workflow converges robustly and reproduces reference equilibrium geometries and energies within tight tolerances.
Most importantly, OOpCCD produces structural parameters that deviate by approximately 0.02 \AA{} (0.01 \AA{}) for bond lengths or less than 1$^\circ$ for bond angles from CCSD(F12c)(T*) (MP2) reference structures. 

\end{abstract}

\maketitle
\section{Introduction}

Geometry optimization is a cornerstone of computational quantum chemistry, enabling the determination of equilibrium structures, transition states, and reaction pathways that govern molecular stability, reactivity, and spectroscopic properties.\cite{schlegel2011geometry, schlegel2003exploring, hratchian2005finding}
Within the Born--Oppenheimer approximation, the identification of chemically-relevant structures such as equilibrium and transition states corresponds to finding stationary points on the potential-energy surface (PES), e.g., by minimizing the energy with respect to nuclear coordinates.\cite{jensen2007introduction}
Access to the first- and second-order derivatives of the electronic energy with respect to nuclear displacements, i.e., the molecular gradient and Hessian, enables structure prediction and characterization of molecular vibrational properties.\cite{helgaker1988analytical}

Modern geometry optimizers combine efficient energy-derivative evaluations with robust optimization algorithms such as quasi-Newton methods and advanced coordinate transformations.\cite{fletcher1987practical, broyden1970convergence}
Internal coordinate representations are particularly advantageous because they reflect chemically intuitive molecular motions and reduce coupling in the approximate Hessian compared to Cartesian coordinates.\cite{pulay1992geometry,bakken2002efficient, fogarasi1992calculation, baker1996generation}
Although numerical gradients obtained via finite-difference schemes are conceptually straightforward, they are significantly more expensive and sensitive to numerical noise and step-size choices.\cite{schlegel2011geometry}
Analytic gradients are therefore preferred in most modern electronic-structure codes because they provide improved efficiency, numerical stability, and scalability.\cite{helgaker1988analytical}

High-accuracy correlated electronic-structure methods, such as coupled-cluster (CC) theory, \cite{cizek_jcp_1966, bartlett2007cc} provide systematic and size-extensive treatments of electron correlation but complicate geometry optimization due to their steep computational scaling and the response equations required for analytic derivatives.\cite{hattig2003geometry, oo-lccd-lambda-equations-pccp-2016}
In conventional post-Hartree--Fock methods, the total derivative of the energy contains contributions from the explicit nuclear-coordinate dependence of the Hamiltonian as well as the implicit response of wavefunction parameters such as orbitals and wavefunction amplitudes.
The implicit response contributions are typically handled through Lagrangian or Z-vector formalisms that transform the gradient evaluation into derivatives of a stationary functional.\cite{helgaker1988analytical,bozkaya2014omp25, bozkaya2016analytic,feng2019implementation}

Orbital-optimized correlated methods provide an attractive alternative framework for handling such response effects.
By variationally optimizing molecular orbitals together with wavefunction amplitudes (which are optimized via projection in CC theory), these methods incorporate orbital relaxation effects directly into the wavefunction and eliminate explicit orbital-response contributions from the analytic gradient expressions.\cite{bozkaya2013analytic, bozkaya2013orbital}
This property often leads to improved numerical stability and better performance for systems exhibiting strong correlation, symmetry breaking, or near-degeneracy effects.\cite{bozkaya2014omp25}

Among reduced-scaling correlated approaches,\cite{pawel-pccp-geminal-review-2022} the pair coupled-cluster doubles (pCCD) model—originally introduced as the antisymmetric product of one-reference orbital geminals (AP1roG)—has emerged as a promising method for describing strong electron correlation.\cite{limacher_2013, boguslawski2014efficient, Stein2014}
By restricting excitations to electron-pair substitutions, pCCD provides a compact seniority-zero wavefunction ansatz that captures important static correlation effects~\cite{pccd-prb-2016} while maintaining a substantially lower computational cost than conventional coupled-cluster approaches.
{In electronic-structure theory, seniority is defined as the number of unpaired electrons in a Slater determinant.\mbox{\cite{gus_seniority,szalay2017seniority}}}
{Thus, a seniority-zero wavefunction contains only electron-paired configurations.}
When combined with orbital optimization,\cite{boguslawski2014efficient,Stein2014,boguslawski2014nonvariational,ps2-ap1rog, piotrus_mol-phys} the method yields a flexible and physically meaningful orbital basis that is particularly well suited for challenging electronic structures such as bond dissociation and strongly correlated molecular systems\cite{pawel_jpca_2014,pawel-pccp-2015,garza2015,pawel-yb2,ola-qit-actinides-pccp-2022} and that provides a convenient basis to interpret electronic excitations and charge-transfer processes in a chemically intuitive way.\cite{pccd-delaram-rsc-adv-2023,pccd-perspective-jpcl-2023,szczuczko2025domain}

Such strong correlation effects frequently arise in extended $\pi$-conjugated systems, including organic semiconductors and donor--acceptor complexes that form the active components of organic photovoltaic (OPV) materials.\cite{takimiya2014pi,su2012organic, zhu2022single}
Accurate modeling of these materials requires reliable descriptions of bond-length alternation, charge-transfer states, and structural relaxation effects that influence exciton dissociation and charge-transport processes.\cite{bredas2009molecular, bredas2014mind, pccd-perspective-jpcl-2023}
Reliable geometry optimization, therefore, plays a crucial role in predicting structural parameters that directly affect electronic couplings, frontier orbital alignment, and energy-level offsets in organic electronic systems.

In this work, we introduce a geometry-optimization engine within the PyBEST software package, \cite{pybest-gpu-jctc-2024, boguslawski2024pybest} enabling analytic nuclear gradient calculations for restricted orbital-optimized (OO)pCCD wavefunctions.
The engine interfaces directly with the \texttt{geomeTRIC} optimizer.\cite{wang2016geometry}
For OOpCCD, the gradient is formulated within a Lagrangian framework and obtained by contracting one- and two-body response density matrices with the first derivatives of molecular integrals.\cite{bozkaya2014omp25, boguslawski2014nonvariational, pccd-ptx}
We benchmark the OOpCCD geometry-optimizer for a test set containing small organic molecules and compare the corresponding bond lengths, bond angles, and dihedral angles to MP2 (analytic gradients) and CCSD(F12c)(T*) (numerical gradients) reference data.

This work is organized as follows. In section~\ref{sec:theory}, we briefly review the evaluation of the OOpCCD molecular gradient using a Lagrangian framework.
Numerical results are presented in section~\ref{sec:results}.
Finally, we conclude in section~\ref{sec:conclusion}.

\section{Geometry optimization}\label{sec:theory}

Let $\mathbf{R}\in\mathbb{R}^{3N}$ denote the Cartesian nuclear coordinates of an $N$-atom system.
The electronic structure method defines a Born--Oppenheimer potential-energy surface (PES) $E(\mathbf{R})$, and an equilibrium geometry corresponds to a stationary point $\mathbf{R}^\star$ of this PES, 
\begin{equation}
\min_{\mathbf{R}}\, E(\mathbf{R}), 
\end{equation}
for which the nuclear forces vanish, $\nabla_{\mathbf{R}} E(\mathbf{R}^\star)=\mathbf{0}$.
In practice, the minimum is located iteratively by generating a sequence of geometries $\{\mathbf{R}^{(k)}\}$ with decreasing energies until standard convergence criteria are satisfied (e.g., thresholds on the maximum and root-mean-square (RMS) force components and on the maximum and RMS displacements).

Although energies and analytic gradients are naturally evaluated in Cartesian coordinates, geometry updates are often performed more efficiently in internal coordinates, which reduce coupling between motions and allow larger, better-conditioned optimization steps than Cartesian displacements.
The \texttt{geomeTRIC} package offers various coordinates (cartesian, primitive internal coordinates, delocalized internal coordinates, etc.).
In this work, the optimization is carried out in the translation--rotation--internal coordinate (TRIC) framework, which augments conventional internal coordinates with explicit collective translations and rotations of molecular subunits, thereby improving robustness for systems containing soft intermolecular degrees of freedom.\cite{wang2016geometry}
At each iteration, the electronic-structure backend supplies the energy and Cartesian gradient $\mathbf{g}^{(k)}=\nabla_{\mathbf{R}}E(\mathbf{R}^{(k)})$, while the optimizer determines a step $\Delta\mathbf{q}$ in the chosen internal coordinates $\mathbf{q}(\mathbf{R})$ and maps it back to a Cartesian displacement using the associated coordinate-transformation machinery.\cite{wang2016geometry} 
The availability of analytic gradients is essential for efficient and numerically stable optimizations, as it avoids the additional cost and noise sensitivity of finite-difference differentiation in each step of the geometry traversal.\cite{hattig2003geometry}
Below, we scrutinize how the energy gradient within the pCCD framework can be efficiently evaluated.
This gradient information is then processed by \texttt{geomeTRIC} to target specific points (minima, transition states) along the reaction coordinate.

\subsection{Analytic gradients: variational and Lagrangian formulation}
Analytic nuclear gradients are first derivatives of the Born--Oppenheimer energy with respect to nuclear coordinates. For a given geometry parameter $x$, the electronic energy may be written as $E\big(x, \boldsymbol{\theta}(x)\big)$, where $\boldsymbol{\theta}$ denotes wave-function parameters (e.g., orbital-rotation parameters and wavefunction amplitudes) that depend implicitly on the molecular geometry.
The total derivative therefore reads,\cite{helgaker1988analytical, yamaguchi2011analytic}
\begin{equation}
\frac{dE}{dx}
=
\frac{\partial E}{\partial x}
+
\sum_p
\frac{\partial E}{\partial \theta_p}\, 
\frac{\partial \theta_p}{\partial x},
\label{eq:chain-rule}
\end{equation}
where we assume that the wavefunction has $p$ geometry-dependent parameters.

If the energy is variational with respect to all parameters, $\partial E/\partial \theta_p = 0$, the response term in Eq.~\eqref{eq:chain-rule} vanishes. In this case, the energy derivative reduces
to the Hellmann--Feynman form,\cite{ feynman1939forces}
\begin{equation}
\frac{dE}{dx}
=
\left\langle \Psi \left| \frac{\partial \hat H}{\partial x} \right| \Psi \right\rangle, 
\label{eq:hf-theorem}
\end{equation}
provided the one-electron basis is independent of $x$.\cite{helgaker1988analytical}
For practical atom-centered Gaussian bases, the basis depends on the nuclear positions, and additional overlap-derivative (Pulay) terms arise even for variational wavefunctions.\cite{pulay1969ab, pulay1980convergence, helgaker1988analytical}

For selected correlated wave-function methods, like coupled-cluster-type models, the energy is not fully variational with respect to the amplitudes and/or the orbitals.
Thus, Eq.~\eqref{eq:chain-rule} contains implicit-response contributions.
Evaluating $\partial\theta_p/\partial x$ explicitly is generally impractical and can be avoided by introducing a Lagrangian (or Z-vector) formulation.\cite{helgaker1988analytical, bozkaya2014omp25, feng2019implementation}
For that purpose, we define the energy Lagrangian
\begin{equation}
\mathcal{L}(x, \boldsymbol{\theta}, \boldsymbol{\lambda})
=
E(x, \boldsymbol{\theta})
+
\boldsymbol{\lambda}^{\mathrm T}\mathbf{c}(x, \boldsymbol{\theta}), 
\label{eq:lagrangian-general}
\end{equation}
where $\mathbf{c}=0$ collects the stationarity conditions of the method (e.g., amplitude equations, orbital-stationarity conditions, and orthonormality constraints) and $\boldsymbol{\lambda}$ are the corresponding Lagrange multipliers (``$\Lambda$''/Z-vector parameters).\cite{handy1984evaluation, helgaker1988analytical, feng2019implementation}
At the stationary point, the conditions
\begin{equation}
\frac{\partial \mathcal{L}}{\partial \boldsymbol{\theta}} = \mathbf{0}, 
\qquad
\frac{\partial \mathcal{L}}{\partial \boldsymbol{\lambda}} = \mathbf{0}
\end{equation}
ensure that all implicit response contributions are accounted for through the multipliers, and the total derivative becomes a simple partial derivative,\cite{helgaker1988analytical, helgaker1989numerically, helgaker1989configuration, helgaker1984second}
\begin{equation}
\frac{dE}{dx}
=
\frac{\partial \mathcal{L}}{\partial x}.
\label{eq:lagrangian-derivative}
\end{equation}
In practice, Eq.~\eqref{eq:lagrangian-derivative} yields gradients as contractions of geometry derivatives of one- and two-electron integrals (and overlap derivatives, giving Pulay terms) with effective one- and two-particle density matrices that include response information.\cite{pulay1969ab, helgaker1988analytical, feng2019implementation}
For orbital-optimized methods, orbital stationarity further simplifies the final expressions by eliminating explicit orbital-response contributions.
In the following, we will describe the derivation of the molecular gradient for a restricted OOpCCD wavefunction.

\subsection{Pair Coupled Cluster Doubles (pCCD/AP1roG)}
\label{sec:pccd-theory}
The pair coupled-cluster doubles (pCCD), also known as the antisymmetric product of one-reference orbital geminals (AP1roG), is a seniority-zero coupled-cluster model in which only electron-pair excitations are included in the cluster operator.\cite{limacher_2013, boguslawski2014efficient, pawel-pccp-geminal-review-2022}
The pCCD wave function is defined through an exponential ansatz 
\begin{equation}
\ket{\Psi_{\rm pCCD}}
=
e^{\Tp}
\ket{\Phi_0}, 
\qquad
\Tp
=
\sum_{i}^{n_{\rm occ}}
\sum_{a}^{n_{\rm virt}}
t_i^{a}\, 
\hat a^{\dagger}_{a}
\hat a^{\dagger}_{\bar a}
\hat a_{\bar i}
\hat a_{i}, 
\label{eq:pccd-wf-theory}
\end{equation}
where $\ket{\Phi_0}$ is a closed-shell reference determinant, $i$ and $a$ denote occupied and virtual spatial orbitals, respectively, and $t_i^{a}$ are the pair amplitudes.
{We emphasize that, in the orbital-optimized pCCD framework, $\ket{\Phi_0}$ is not the conventional Hartree--Fock determinant.}
{Instead, it is defined in the optimized orbital basis and is therefore a system-adapted reference determinant within the seniority-zero model.}
{This point is closely related to other seniority-based approaches, such as CCD0, where the restricted excitation manifold defines a special reference and correlation space; as discussed by Bulik and co-workers, adding back additional cluster-operator terms does not necessarily lead to systematic improvements for strongly correlated systems.\mbox{~\cite{bulik2015can}}}
Spin degrees of freedom are indicated by barred and unbarred indices, namely, $i$ for $\alpha$ spin and $\bar i$ for $\beta$ spin, respectively.
The operator $\Tp$ promotes electron pairs from occupied (occ) to virtual (virt) orbitals while preserving seniority (the number of unpaired electrons).~\cite{Stein2014, ps2-ap1rog, boguslawski2014nonvariational, piotrus_mol-phys, pccd-prb-2016, gauss1991coupled}
We should note that by restricting the excitation manifold to paired double excitations, pCCD captures a substantial portion of static (nondynamic) correlation effects at a computational cost comparable to mean-field methods.\cite{limacher_2013, boguslawski2014nonvariational}
However, dynamic correlation arising from broken-pair excitations is not described and must be recovered through \textit{a posteriori} corrections, such as perturbation theory, linearized coupled-cluster corrections, or tailored CC approaches.\cite{frozen-pccd, piotrus_pt2, pccd-ptx, tailored-pccd-jctc-2022}
Size-consistency of pCCD is recovered when the orbitals are optimized variationally.
In OOpCCD, the energy is made stationary with respect to both the pair amplitudes and orbital rotations.\cite{Stein2014, boguslawski2014nonvariational,ps2-ap1rog}
Orbital optimization significantly improves equilibrium geometries and spectroscopic constants and simplifies analytic derivative theory by eliminating explicit orbital-response contributions from the final gradient expression.\cite{bozkaya2014omp25}
{In fact, orbital optimization is essential for seniority-zero methods, since the quality of the pCCD/AP1roG wavefunction strongly depends on the molecular orbital basis.}
{Without orbital optimization, the seniority-zero restriction can lead to an unbalanced and qualitatively poor description of electron correlation, especially for strongly correlated systems, and a lack of size-consistency.}
Therefore, we will focus solely on the OOpCCD model and its performance in yielding reliable molecular geometries.

{Importantly, the pCCD response one- and two-particle reduced density matrices (1-RDM and 2-RDM) admit compact expressions that retain a pair structure.}\cite{jahani2026ionization}
{These response density matrices enable an efficient evaluation of analytic nuclear gradients, analogous to standard coupled-cluster derivative theory, but without the need to solve additional orbital-response or amplitude-response equations.}\cite{helgaker1988analytical, behjou2025electron, jahani2025simple}


OOpCCD evaluates analytic nuclear gradients within the standard Lagrangian (Z-vector) framework used for non-variational correlated methods.\cite{helgaker1988analytical, boguslawski2014nonvariational, bozkaya2014omp25, Stein2014, handy1984evaluation}
In this approach, all implicit parameter-response contributions (cluster amplitudes, orbital rotations, and the $\lambda$ amplitudes of the pCCD $\Lambda$ equations) are accounted for through stationarity of the energy Lagrangian, so that the total derivative of the energy with respect to a
nuclear coordinate can be evaluated as a simple \emph{partial} derivative of $\mathcal{L}$ at the stationary point.\cite{helgaker1988analytical}
Specifically, the OOpCCD energy Lagrangian reads~\cite{Stein2014, boguslawski2014nonvariational, ps2-ap1rog, jahani2026ionization}
\begin{align}
\mathcal{L} 
    &=
        \langle{\Phi_0} | {e^{-\Tp} e^{\hat \kappa} \hat H e^{-\hat \kappa} e^{\Tp}}  | {\Phi_0} \rangle \nonumber \\
    &\phantom{==}+ \sum_{ia}
        \lambda_i^a\, 
        \langle{\Phi_{i\bar i}^{a\bar a}}   | 
       {e^{-\Tp} e^{\hat \kappa} \hat H e^{-\hat \kappa} e^{\Tp}}
         | {\Phi_0} \rangle \nonumber \\
    &=
\left\langle \Phi_0 \left| (1+\hat \Lambda)\, e^{-\Tp} e^{\hat \kappa} \hat H e^{-\hat \kappa} e^{\Tp} \right| \Phi_0 \right\rangle,
\label{eq:oopccd-lagrangian}
\end{align}
where $\{\lambda_i^a\}$ are Lagrange multipliers
with the pCCD de-excitation operator
\begin{equation}
\hat \Lambda = \sum\limits_{ia} \lambda_i^a\, \hat a_i^\dagger \hat a_{\bar{i}}^\dagger \hat a_{\bar{a}} \hat a_a,
\label{eq:oopccd-Lambda-op}
\end{equation}
{and the molecular Hamiltonian $\hat H$ is written in second-quantized form as}
\begin{equation}
\hat H =
\sum_{pq,\sigma} h_{pq}\,
\hat a_{p_\sigma}^{\dagger}\hat a_{q_\sigma}
+
\frac{1}{2}
\sum_{pqrs,\sigma\tau}
g_{pqrs}\,
\hat a_{p_\sigma}^{\dagger}
\hat a_{q_\tau}^{\dagger}
\hat a_{s_\tau}
\hat a_{r_\sigma}
+
V_{\rm nn},
\label{eq:coulomb-hamiltonian}
\end{equation}
{where $h_{pq}$ and $g_{pqrs}$ are the one- and two-electron integrals, respectively, in a given orbital basis, $V_{\rm nn}$ denotes the nuclear--nuclear repulsion energy, and $\sigma,\tau$ are spin labels.}
{This Hamiltonian is later transformed to the geometry-dependent orthonormal MO basis used in the gradient derivation.}
{The operator} $\hat \kappa$ is the generator for orbital rotations.\cite{feng2019implementation, levchenko2005analytic}
For (real) spatial orbitals, we rewrite $\hat \kappa$ as
\begin{equation}
\hat \kappa = \sum_{p > q} \kappa_{pq} \big( \hat E_{pq} - \hat E_{qp} \big), 
\label{eq:oopccd-kappa}
\end{equation}
where $\hat E_{pq} = \hat a_p^\dagger \hat a_q + \hat {a}_{\bar p}^\dagger \hat {a}_{\bar q}$ represents the
singlet excitation operator, with $p$ and $q$ running over all active orbitals, and
$|\Phi^{{a}\bar{a}}_{{i}\bar{i}}\rangle = \hat a^\dagger_a \hat a^\dagger_{\bar{a}} \hat a_{\bar{i}} \hat a_i |\Phi_0\rangle$.

The stationary conditions with respect to the pCCD parameters yield the usual pair-amplitude (geminal-coefficient) equations and the associated $\Lambda$ equations.\cite{boguslawski2014nonvariational, Stein2014, ps2-ap1rog, jahani2026ionization}
The stationary condition with respect to the $\lambda$ amplitudes leads to the pair-doubles amplitude equations,
\begin{equation}
\frac{\partial \mathcal{L}}{\partial \lambda_i^a} \Big\vert_{\kappa=0}
= \langle \Phi_{i{\bar{i}}}^{a{\bar{a}}} | e^{-\Tp} \hat H e^{\Tp} | \Phi_0 \rangle = 0, 
\label{eq:oopccd-amp-eq}
\end{equation}
while the corresponding pCCD $\Lambda$-equations are obtained from
\begin{align}
\left.\frac{\partial \mathcal{L}}{\partial t_i^{a}}\right|_{\kappa=0}
    &= \langle \Phi_0 | e^{-\Tp} \hat H e^{\Tp}\, \hat a_a^\dagger \hat a_{\bar a}^\dagger \hat a_{\bar i} \hat a_i | \Phi_0 \rangle \nonumber \\
    &\phantom{==}+ \sum_{j b} \lambda_j^b
        \langle \Phi_{j\bar j}^{\, b\bar b} | e^{-\Tp} \hat H e^{\Tp}\, \hat a_a^\dagger \hat a_{\bar a}^\dagger \hat a_{\bar i} \hat a_i | \Phi_0 \rangle  \nonumber \\
    &= 0.
\label{eq:oopccd-lambda-eq}
\end{align}
For the orbital gradient, stationarity requires
\begin{align}
\frac{\partial \mathcal{L}}{\partial \kappa_{pq}} \Big\vert_{\kappa=0}
    &= \langle \Phi_0 | (1+\hat \Lambda)\, 
        e^{-\Tp}
        \big[ (\hat E_{pq} - \hat E_{qp}), \hat H \big]
        e^{\Tp}
        | \Phi_0 \rangle \nonumber \\
    &= 0.
\label{eq:oopccd-orb-stationarity}
\end{align}

In practice, orbital optimization enforces Eq.~\eqref{eq:oopccd-orb-stationarity}, 
which is a key simplification for first-derivative theory:
{once the orbitals satisfy the Lagrangian orbital-stationarity condition},
explicit orbital-response terms do not appear in the final analytic gradient expression.\cite{bozkaya2014omp25}

For efficient evaluation, the OOpCCD analytic gradient can be expressed using response one- and two-particle reduced density matrices (RDMs), contracted with the corresponding first derivatives of the one- and two-electron integrals. 
For pCCD, the response 1-RDM is diagonal, 
\begin{equation}
\gamma^p_p
=
\langle \Phi_0 | (1+\hat \Lambda)\, e^{-\Tp} \hat a_p^\dagger \hat a_p e^{\Tp} | \Phi_0 \rangle, 
\label{eq:oopccd-1rdm}
\end{equation}
and the response 2-RDM, determined from
\begin{equation}
\Gamma^{pq}_{rs}
=
\langle \Phi_0 | (1+\hat \Lambda)\, e^{-\Tp} \hat a_p^\dagger \hat a_q^\dagger \hat a_s \hat a_r e^{\Tp} | \Phi_0 \rangle, 
\label{eq:oopccd-2rdm}
\end{equation}

{For completeness, we provide the explicit non-zero response one- and two-particle reduced density matrix elements entering the OOpCCD gradient expression.}
{Since we work with a seniority-zero wave function and molecular orbitals, the only non-zero blocks of the spin-integrated response 2-RDM are $\Gamma^{pq}_{pq}$, $\Gamma^{p\bar q}_{p\bar q}$, $\Gamma^{p\bar p}_{q\bar q}$.
}
{Furthermore, the spin-integrated 2-RDM blocks satisfy}
\begin{equation}
\Gamma^{pq}_{pq}
=
\Gamma^{p\bar q}_{p\bar q},
\qquad
\Gamma^{pq}_{pq}
=
\Gamma^{qp}_{qp},
\qquad
\Gamma^{p\bar p}_{p\bar p}
=
\gamma_p^p,
\end{equation}
{and, in the case of (ROOpCCD) response RDMs within the Lagrangian formulation,}
\begin{equation}
\Gamma^{p\bar p}_{q\bar q}
\neq
\Gamma^{q\bar q}_{p\bar p}.
\end{equation}
{Thus, only two independent blocks of the full response 2-RDM need to be stored in the molecular-orbital basis.}
{Using occupied indices $i,j$ and virtual indices $a,b,c$, the diagonal response 1-RDM elements are}
\begin{align}
\gamma^{i}_{i}
    &= 1 - \sum_c t_i^c \lambda_i^c,
&
\gamma^{a}_{a}
    &= \sum_k t_k^a \lambda_k^a.
\label{eq:rdms}
\end{align}
{The non-zero response 2-RDM elements are}
\begin{align}
\Gamma^{i\bar{j}}_{i\bar{j}}
=
\Gamma_{ij}^{ij}
&=
1
-
\sum_c \lambda_i^c t_i^c
-
\sum_c \lambda_j^c t_j^c,
\qquad i\neq j,
\nonumber\\
\Gamma^{i\bar{a}}_{i\bar{a}}
=
\Gamma^{a\bar{i}}_{a\bar{i}}
=
\Gamma^{ia}_{ia}
=
\Gamma^{ai}_{ai}
&=
\sum_k \lambda_k^a t_k^a
-
\lambda_i^a t_i^a,
\nonumber\\
\Gamma^{i\bar{i}}_{j\bar{j}}
&=
\sum_c \lambda_j^c t_i^c
+
\delta_{ij}
\left(
1
-
2\sum_c \lambda_i^c t_i^c
\right),
\nonumber\\
\Gamma^{i\bar{i}}_{a\bar{a}}
&=
t_i^a
+
2\lambda_i^a t_i^a t_i^a
-
2\sum_k \lambda_k^a t_k^a t_i^a
\nonumber\\
&\phantom{-}-
2\sum_c \lambda_i^c t_i^c t_i^a
+
\sum_{kc}\lambda_k^c t_i^c t_k^a,
\nonumber\\
\Gamma^{a\bar{a}}_{i\bar{i}}
&=
\lambda_i^a,
\nonumber\\
\Gamma^{a\bar{a}}_{b\bar{b}}
&=
\sum_k \lambda_k^a t_k^b.
\label{eq:rdms_all}
\end{align}
{All remaining response 2-RDM elements vanish due to the seniority-zero restriction of the pCCD wave function.}
These sparse, pair-structured RDMs are a defining feature of seniority-zero models and are the reason why OOpCCD gradients can be evaluated at substantially reduced cost compared to general coupled-cluster gradients.



For OOpCCD, the total derivative with respect to the nuclear-coordinate vector $\bm{x}$ can be written as
\begin{align}
\frac{d E_{\rm OOpCCD}}{d \bm{x}}
 &=
\frac{\partial \mathcal{L}}{\partial \bm{x}}
+ \sum_{\mu} \frac{\partial \mathcal{L}}{\partial t_{\mu}} \frac{\partial t_{\mu}}{\partial \bm{x}}
+ \sum_{\mu} \frac{\partial \mathcal{L}}{\partial \lambda_{\mu}} \frac{\partial \lambda_{\mu}}{\partial \bm{x}}
+ \sum_{pq} \frac{\partial \mathcal{L}}{\partial \kappa_{pq}} \frac{\partial \kappa_{pq}}{\partial \bm{x}} \nonumber \\
 &= \frac{\partial \mathcal{L}}{\partial \bm{x}} \nonumber \\
 &=
\left\langle \Phi_0 \left| (1+\hat \Lambda)\, e^{-\Tp} e^{\hat \kappa}
\frac{\partial H}{\partial \bm x}
e^{-\hat \kappa} e^{\Tp} \right| \Phi_0 \right\rangle .
\label{eq:oopccd-total-deriv}
\end{align}
where the first reduction follows from stationarity conditions of the energy Lagrangian; $\partial\mathcal{L}/\partial t_\mu=0$ (amplitude equations of Eq.~\eqref{eq:oopccd-amp-eq}), $\partial\mathcal{L}/\partial \lambda_\mu=0$ ($\Lambda$ equations of Eq.~\eqref{eq:oopccd-lambda-eq}), and $\partial\mathcal{L}/\partial \kappa_{pq}=0$ (orbital gradient of Eq.~\eqref{eq:oopccd-orb-stationarity}).
The second-quantized form of the molecular Hamiltonian $\hat H$ depends on the nuclear coordinates.
It is custom to express the Hamiltonian in a new geometry-dependent basis, known as the \emph{orthonormal MO} (OMO) basis, which is obtained from the original basis via Löwdin symmetric orthonormalization.\cite{helgaker1988analytical, bozkaya2014omp25}
In this OMO basis, the molecular Hamiltonian,
\begin{align}
    \hat H(\bm x) &=   
        \sum_{r_\sigma s_\sigma} \tilde h_{r_\sigma s_\sigma} \hat a_{r_\sigma}^\dagger \hat a_{s_\sigma} \nonumber \\
    &\phantom{==}+   \frac{1}{2} \sum_{r_\sigma s_\tau t_\sigma u_\tau} \tilde g_{r_\sigma s_\tau t_\sigma u_\tau} \hat a^\dagger_{r_\sigma} \hat a_{s_\tau}^\dagger \hat a_{u_\tau} \hat a_{t_\sigma}
    + V_{nn}
\end{align}
contains the transformed integrals (dropping the explicit spin-dependence)
\begin{equation}
\tilde{h}_{pq}(\bm x)
=
\sum_{p^\prime q^\prime} h_{p^\prime q^\prime}(\bm x)
(S^{-\nicefrac{1}{2}})_{pp^\prime}(\bm x)
(S^{-\nicefrac{1}{2}})_{qq^\prime}(\bm x)
\label{eq:lowdin-h}
\end{equation}
and (the indices are ordered according to Physicist's notation)
\begin{align}
\tilde{g}_{pqrs}
    =   &\sum_{p^\prime q^\prime r^\prime s^\prime} g_{p^\prime q^\prime r^\prime s^\prime}(\bm x)
        (S^{-\nicefrac{1}{2}})_{pp^\prime}(\bm x)
        (S^{-\nicefrac{1}{2}})_{qq^\prime}(\bm x) \nonumber \\
    &   (S^{-\nicefrac{1}{2}})_{rr^\prime}(\bm x)
        (S^{-\nicefrac{1}{2}})_{ss^\prime}(\bm x),
\label{eq:lowdin-g}
\end{align}
where $S$ is the overlap matrix of the AO basis evaluated at point $\bm x$.
Evaluating first derivatives at a reference geometry $\bm x_0$, one obtains the standard symmetric-connection expressions for the derivatives of the integrals expressed in an orthonormal-basis,\cite{bozkaya2014omp25}
\begin{equation}
\tilde{h}^{(x)}_{pq}
=
h^{(x)}_{pq}
-\frac{1}{2}\sum_{t}
\Big(
h_{tq}\, S^{(x)}_{tp}
+
h_{pt}\, S^{(x)}_{tq}
\Big), 
\label{eq:lowdin-h-deriv}
\end{equation}
\begin{equation}
\tilde{g}^{(x)}_{pqrs}
=
g^{(x)}_{pqrs}
-\frac{1}{2}\sum_{t}
\Big(
g_{tqrs}\, S^{(x)}_{tp}
+
g_{ptrs}\, S^{(x)}_{tq}
+
g_{pqts}\, S^{(x)}_{tr}
+
g_{pqrt}\, S^{(x)}_{ts}
\Big).
\label{eq:lowdin-g-deriv}
\end{equation}
Here, $h^{(x)}$, $g^{(x)}$, and $S^{(x)}$ denote the first partial derivatives of the one-electron, two-electron, and overlap integrals with respect to some nuclear coordinate $x$, and the indices $p, q, r, s, t$ label molecular orbitals.

Starting from Eq.~\eqref{eq:oopccd-total-deriv} and evaluating the partial derivative for the current set of orbitals and nuclear coordinates, i.e., at $\kappa=0$ and $\bm x = \bm x_0$, the gradient of the energy can be determined from
\begin{widetext}
\begin{align}
\frac{dE_{\rm OOpCCD}}{dx}
&=
\sum_{pq, \sigma} \tilde h^{(x)}_{pq} \left\langle \Phi_0 \left| (1+\hat \Lambda)\, e^{-\Tp} \hat a_{p_\sigma}^\dagger \hat a_{q_\sigma} e^{\Tp} \right| \Phi_0 \right\rangle \nonumber \\
&+
\frac12\sum_{pqrs,\sigma \tau} \tilde g^{(x)}_{pqrs} \left\langle \Phi_0 \left| (1+\hat \Lambda)\, e^{-\Tp} \hat a_{p_\sigma}^\dagger \hat a_{q_\tau}^\dagger \hat a_{s_\tau} \hat a_{r_\sigma} e^{\Tp} \right| \Phi_0 \right\rangle 
+ \frac{dV_{\rm nn}}{dx},
\end{align}
\end{widetext}
where we dropped the spin-dependence of the integrals as we work in a restricted spatial orbital basis.
Substituting the response 1- and 2-RDMs from eqs.~\eqref{eq:oopccd-1rdm} and \eqref{eq:oopccd-2rdm} into the above equation, the energy gradient can be written in compact form (again using restricted spatial orbitals)
\begin{align}\label{eq:oogradient}
\frac{dE_{\rm OOpCCD}}{dx}
    &= 2 \sum_{p}\gamma_p^p\, \tilde h^{(x)}_{pp} + \sum_{pq} (2\tilde g^{(x)}_{pqpq}-\tilde g^{(x)}_{pqqp})\Gamma_{pq}^{pq} \nonumber \\
    &\phantom{==}  + {\sum_{pq} \tilde g^{(x)}_{ppq{q}}\tilde \Gamma_{p\bar p}^{q \bar q}} +
    \frac{dV_{\rm nn}}{dx},
\end{align}
where we introduced the symmetrized 2-RDM block $\tilde \Gamma_{p\bar p}^{q \bar q} = \frac12 (\Gamma_{p\bar p}^{q \bar q} + \Gamma_{q\bar q}^{p \bar p})$.
{For a variationally-optimized seniority-zero wave function, the corresponding 2-RDM elements are identical by symmetry.}
{However, in the present Lagrangian formulation, the response 2-RDM contains $\Lambda$-amplitude contributions, and the pair-transfer response blocks are not necessarily symmetric term by term.}
{The symmetrized form allows us to evaluate the (orbital) gradient more quickly.}
{To be consistent with our OOpCCD implementation when evaluating the generalized Fock matrix (\textit{vide infra}), this notation has been adapted to nuclear gradient theory for reasons of consistency (as introduced in the above equation).}
{Eq.~\mbox{\eqref{eq:oogradient}}} can be further rewritten using the generalized Fock matrix of OOpCCD,~\cite{jahani2026ionization}

\begin{align}
    F_{qp} 
    &= h_{pq} \gamma_q^q 
        + {\sum_u} (2g_{puqu} - g_{puuq})\Gamma_{qu}^{qu} 
        + {\sum_u} g_{pquu} \tilde{\Gamma}_{u\bar{u}}^{q\bar{q}}.
    \label{eq:gfm}
\end{align}


By inserting the expressions for $\tilde h^{(x)}_{pq}$ of Eq.~\eqref{eq:lowdin-h-deriv} and $\tilde g^{(x)}_{pqrs}$ of Eq.~\eqref{eq:lowdin-g-deriv}, the OOpCCD energy gradient can be evaluated from (again, using restricted spatial orbitals)
\begin{align}
\frac{dE_{\rm OOpCCD}}{dx}
    &=  2 \sum_{p}\gamma_p^p\, h^{(x)}_{pp} + \sum_{pq} (2 g^{(x)}_{pqpq}-g^{(x)}_{pqqp})\Gamma_{pq}^{pq} \nonumber \\
    &\phantom{==} + \sum_{pq} g^{(x)}_{ppqq}\tilde \Gamma_{p\bar p}^{q \bar q} - 2\sum_{pq} F_{pq}\, S_{qp}^{(x)} +
\frac{dV_{\rm nn}}{dx}.
    \label{eq:oopccd-gradient}
\end{align}
In the above equation, $h^{(x)}_{pq}$ and $g^{(x)}_{pqrs}$ again denote the first partial derivatives of the one- and two-electron integrals with respect to the nuclear coordinate $x$, while $S^{(x)}_{pq}$ is the derivative of the overlap matrix.
The first three terms correspond to the Hellmann--Feynman contributions arising from the explicit geometry dependence of the Hamiltonian integrals.
The third term represents the Pulay correction originating from the geometry dependence of the atomic orbital basis functions, and $V_{\rm nn}$ is the nuclear–nuclear repulsion energy.

Analytic gradient implementations typically follow a common structure: the energy derivative is expressed as contractions between derivative integrals and effective one- and two-particle density matrices, together with additional contributions arising from the geometry dependence of the
basis functions (Pulay terms).\cite{pulay1969ab, pulay1980convergence, helgaker1988analytical, gauss1991analytic}
Because derivative integrals are naturally evaluated in the atomic-orbital (AO) basis whereas correlated response quantities are most conveniently expressed in the molecular-orbital (MO) representation, efficient implementations rely on consistent AO--MO transformations of
density matrices and related quantities.\cite{helgaker1988analytical, szabo_book}

Due to the seniority-zero restriction of the CC ansatz, the corresponding (response) 1- and 2-RDMs have a particularly simple and sparse form.
As a consequence, it is computationally more efficient to evaluate the OOpCCD energy gradient in the MO basis.
Although we have to perform a 4-index transformation of the first-order derivatives of the electron-repulsion integrals, the 2-RDM blocks remain sparse and can be stored as (separate) two-dimensional arrays.
Moving to the AO basis, the sparse MO-based 2-RDM must be back-transformed into the full four-dimensional 2-RDM counterpart, which requires $\mathcal{O}(N^4)$ storage.
On the other hand, the one-electron term and the generalized Fock matrix term can be easily back-transformed into the AO picture, losing again the sparsity of MO-based 1-RDM (which becomes a two-dimensional object in the AO basis).
Since the evaluation of the two-body terms is the most cost-intensive step, we have implemented the two-electron terms in the evaluation of the OOpCCD energy gradient in the MO basis.
{Compared with conventional CCSD, the seniority-zero restriction in OOpCCD substantially reduces the storage requirements and the cost of contractions involving the response 2-RDM, since only pair-structured two-dimensional blocks are required. The evaluation of the generalized Fock matrix scales as $\mathcal{O}(N^3)$, while the overall gradient step can still involve higher-scaling operations, including derivative-integral generation and integral transformations.}

\section{Computational Details}\label{sec:Computational_Details}

All implementations and calculations were carried out with a development version of the \textsc{PyBEST} program package v2.2.0.dev0,\cite{pybest-paper-2021,pybest-gpu-jctc-2024,boguslawski2024pybest} in which we implemented a geometry-optimization engine interfaced to the \texttt{geomeTRIC} package.\cite{wang2016geometry}
Unless stated otherwise, geometry optimizations were performed in the translation--rotation--internal coordinate (TRIC) framework as implemented in \texttt{geomeTRIC}, using its default convergence thresholds.
Analytic nuclear gradients were evaluated at the restricted Hartree--Fock (RHF) and orbital-optimized pair coupled-cluster doubles (OOpCCD/AP1roG) levels of theory.
For RHF, standard analytic gradients were used.
For OOpCCD, analytic gradients were evaluated within the Lagrangian framework described in Section~\ref{sec:theory}, employing response one- and two-particle reduced density matrices and the corresponding generalized Fock contributions.
{In addition to the diatomic validation tests, we verified the OOpCCD analytic gradients for selected polyatomic molecules by comparing them with central finite-difference numerical gradients.}
{For all tested Cartesian components, the analytic and numerical gradients agreed within the expected finite-difference accuracy, confirming the correctness of the implementation also for polyatomic systems.}
To maximize efficiency and minimize storage, we implemented a mixed-AO-MO analytic gradient evaluation, where the one-body terms are evaluated in the AO basis, while the two-body contributions are computed in the MO basis (due to the sparsity of the response 2-RDMs).
The benchmark calculations were carried out with Dunning’s correlation-consistent basis sets,\cite{cc-pvdz-h-b-ne-jcp-1989} cc-pVDZ and cc-pVTZ.
No frozen-core approximation was employed.
The test set comprises {six} diatomic molecules (BN, C$_2$, CN$^+$, CO, F$_2$, and N$_2$), {14} small closed-shell organic molecules,~\cite{crisci2025energies} and {three} selected transition-state {examples}.~\cite{asgeirsson2021nudged}
{This validation set spans different bonding situations, including single, double, conjugated, cyclic, and aromatic systems, as well as first-order saddle points.}
For the diatomic systems, reference equilibrium bond lengths were also obtained from one-dimensional potential-energy scans.
Single-point energies were computed for several bond distances in the vicinity of the minimum and fitted to a third-order polynomial.
The corresponding equilibrium bond length was then extracted from the fitted curve.
For the molecular benchmark set, equilibrium geometries optimized at the OOpCCD/cc-pVDZ level were compared with reference MP2 and CCSD(F12c)(T*) geometries taken from the literature.\cite{crisci2025energies}
For reasons of comparison, a cc-pVDZ basis was employed, without any frozen-core approximation.
Bond lengths, bond angles, and dihedral angles were extracted from the final optimized structures and analyzed statistically with respect to the reference data (all data is collected in the SI).
Transition-state optimizations were performed with the same \textsc{PyBEST}--\texttt{geomeTRIC} interface.
Initial transition-state guesses were generated from chemically motivated structures.
Specifically, the OOpCCD TS search had to be accelerated by starting with the RHF-optimized TS structure.
The character of the converged stationary points was verified by numerical Hessian calculations, and first-order saddle points were identified by the presence of a single imaginary vibrational frequency.
In all OOpCCD calculations, tight convergence thresholds have been applied, namely,  $10^{-12}$ for the pCCD amplitude residual, $10^{-8}$ for the pCCD total energy, $10^{-5}$ for the maximum element (as absolute value) of the pCCD orbital gradient, and $10^{-4}$ for the pCCD orbital gradient norm.
{In the SI, we collect selected wall times for the OOpCCD gradient evaluations across the benchmark set.
The timings increase with system size and basis-set dimension, with the dominant cost arising from the OOpCCD amplitude solver, orbital optimization, and integral transformations.
The actual gradient evaluation (including the calculation of the derivative integrals) amounts to 10 to 25\% of the geometry optimization step.
The reported timings, therefore, support the reduced-cost character of OOpCCD and show that analytic-gradient geometry optimizations are computationally inexpensive.}

\section{Validation and Results}\label{sec:results}

\begin{figure*}[!t]
\centering
\includegraphics[width=0.8\textwidth]{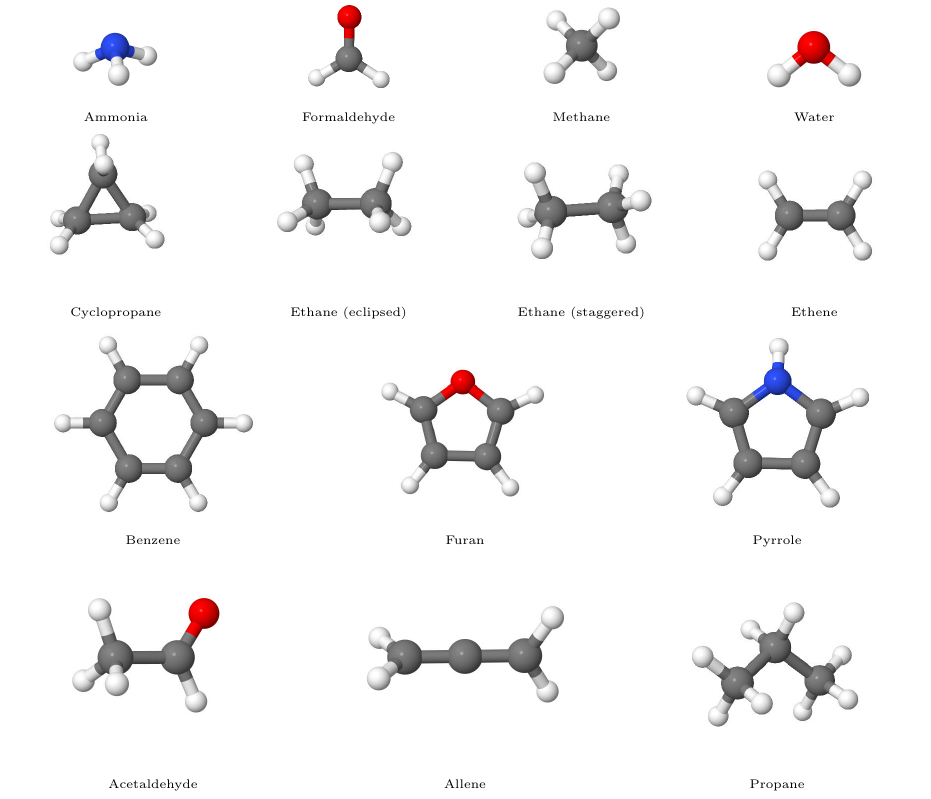}
\caption{Molecular test set used for benchmarking the performance of the analytic-gradient-driven geometry optimization engine within OOpCCD.
The molecular structures are taken from Ref.~\citenum{crisci2025energies} and re-optimized for OOpCCD.
All molecular structures are visualized using the Jmol package.\cite{jmol}}
\label{fig:validation-molecules}
\end{figure*}

We evaluate the numerical reliability and robustness of the OOpCCD molecular structure-optimization engine using a diverse validation set selected to probe both the correctness of the analytic energy gradients and the performance of the optimization procedure across different bonding environments.

Our benchmark set is organized into three categories.
First, a series of diatomic molecules (BN, F$_2$, N$_2$, CN$^+$, CO, and C$_2$) is used to verify the correctness of equilibrium bond lengths obtained from analytic gradient-driven optimizations.
Specifically, their equilibrium bond lengths can be straightforwardly determined from a numerical fit of the PES obtained for points located around the anticipated equilibrium bond lengths.
Second, a collection of small closed-shell organic molecules is considered to assess the performance of the optimizer for typical covalent bonding motifs, including single, double, and conjugated bonds.
For this test set, highly accurate CCSD(F12)(T*) reference data is available, which allows us to assess how well OOpCCD reproduces structural parameters despite the lack of dynamical correlation.
Finally, we probe the OOpCCD molecular structure-optimization engine for targeting transition states along a reaction coordinate.
For that purpose, we select a simple example containing only main-group elements, which is taken from Ref.~\citenum{asgeirsson2021nudged}.

Taken together, these systems provide a compact yet representative assessment of (i) the numerical correctness of the analytic OOpCCD gradients, (ii) the accuracy of optimized equilibrium structures, and (iii) the stability of the TRIC-based optimization workflow to target optimized ground- and transition-state molecular structures.
The second validation set is illustrated in Figure~\ref{fig:validation-molecules}, spanning diatomic, acyclic, and cyclic molecules, while the third test case is visualized in Figure~\ref{fig:pes-ts-search}.

\begin{table}[t]
\caption{Equilibrium bond lengths and total energies from numerical fits to the PES compared to analytic geometry optimizations for different basis sets.
For the PES-fit data, $r_e$ and $E(r_e)$ are obtained by fitting the scanned energies to a 3rd-order polynomial and locating the minimum on the fitted curve with $\Delta x = 0.01$ \AA{} (see SI for more details).
Analytic results report the optimized equilibrium energy $E_e$ and bond length $r_e$ from direct geometry optimization.
}
\label{tab:poly_fit}
\centering
\begin{tabular}{l l rr rr}
\toprule
Molecule & Basis set &
\multicolumn{2}{c}{numeric fit} &
\multicolumn{2}{c}{analytic gradients} \\
\cmidrule(lr){3-4} \cmidrule(lr){5-6}
 &  & $E(r_e)$ [E$_h$] & $r_e$ [\AA]
 & $E_e$ [E$_h$] & $r_e$ [\AA] \\
\midrule
BN     & cc-pVDZ &  -79.029999 & 1.2688 &  -79.029999 & 1.2688 \\
       & cc-pVTZ &  -79.065094 & 1.2582 &  -79.065094 & 1.2582 \\
C$_2$  & cc-pVDZ &  -75.549523 & 1.2387 &  -75.549522 & 1.2387 \\
       & cc-pVTZ &  -75.582371 & 1.2213 &  -75.582369 & 1.2213 \\
CN$^+$ & cc-pVDZ &  -91.803418 & 1.1657 &  -91.803774 & 1.1630 \\
       & cc-pVTZ &  -91.842591 & 1.1499 &  -91.842591 & 1.1499 \\
CO     & cc-pVDZ & -112.855529 & 1.1231 & -112.855529 & 1.1231 \\
       & cc-pVTZ & -112.911671 & 1.1156 & -112.911671 & 1.1156 \\
F$_2$  & cc-pVDZ & -198.855477 & 1.5186 & -198.857066 & 1.5190 \\
       & cc-pVTZ & -198.949747 & 1.4624 & -198.949747 & 1.4624 \\
N$_2$  & cc-pVDZ & -109.062713 & 1.1016 & -109.062706 & 1.1016 \\
       & cc-pVTZ & -109.127740 & 1.0867 & -109.127740 & 1.0867 \\
\bottomrule
\end{tabular}
\end{table}

\subsection{Diatomic equilibrium bond lengths: PES fits vs.\ analytic optimizations}
\label{sec:diatomic-bench}

The diatomic subset provides a direct and sensitive validation of the analytic gradient implementation and the optimization algorithm.
In particular, because the equilibrium geometry of a diatomic molecule corresponds to the minimum of a one-dimensional potential-energy curve, the optimized bond length obtained from analytic gradient calculations should coincide with the minimum of the fitted PES.

Table~\ref{tab:poly_fit} compares equilibrium bond lengths obtained from numeric PES fits with those determined from direct analytic geometry optimizations using our OOpCCD molecular structure optimization engine.
For that purpose, we determined the equilibrium bond lengths from one-dimensional energy scans, followed by polynomial fitting.
For each molecule and basis set, single-point energies were computed at five bond lengths in a narrow region around the minimum, and the resulting $E(r)$ values were least-squares fitted to an 3rd-order polynomial.
The equilibrium distance $r_e$ was then obtained by minimizing the fitted
polynomial, and the corresponding energy $E(r_e)$ was evaluated at the minimum
of the fitted curve.

The results demonstrate excellent agreement between the minima obtained from polynomial PES fitting and those obtained from direct analytic geometry optimizations.
For most systems and basis sets, both the equilibrium energies and bond lengths coincide within the reported numerical precision, indicating that the analytic gradient implementation accurately identifies the stationary point of the potential-energy surface.

\begin{table*}[!t]
\caption{Average, maximum, and median deviations in the structural parameters for the optimized geometries relative to MP2 reference data.
    Reported are the deviations in bond lengths [\AA] and bond angles [$^\circ$] for each molecule (OOpCCD--MP2).
    The ME, MAE, SD, SD-pop, RMSE, and MPE values are determined for each deviation and structural parameter and explained below.
    The MP2 columns list the corresponding reference values (obtained with analytic gradients) for the parameter at which the maximum deviation occurs.
    In both calculations, a cc-pVDZ basis set was applied.
    A similar analysis is performed for CCSD(F12c)(T*) reference data, which is summarized in the SI.
    }
\label{tab:geom-errors}
\centering
\setlength{\tabcolsep}{5pt}
\begin{tabular}{l rr rr rr rr}
\toprule
    & \multicolumn{6}{c}{OOpCCD--MP2} 
    & \multicolumn{2}{c}{MP2\cite{crisci2025energies}}
    \\
    & \multicolumn{2}{c}{average} 
    & \multicolumn{2}{c}{maximum} 
    & \multicolumn{2}{c}{median}
    & \multicolumn{2}{c}{maximum} 
    \\
\cmidrule(lr){2-3}
\cmidrule(lr){4-5}
\cmidrule(lr){6-7}
\cmidrule(lr){8-9}
Molecule 
    & $\Delta R$ [\AA] & $\Delta \angle$ [$^\circ$] 
    & $\Delta R$ [\AA] & $\Delta \angle$ [$^\circ$] 
    & $\Delta R$ [\AA] & $\Delta \angle$ [$^\circ$] 
    & $R$ [\AA] & $\angle$ [$^\circ$] 
    \\
\midrule
acetaldehyde       &  0.0013 &  0.0014&  0.0124 & 1.2065 &  0.0041 & -0.0486 & 1.5096 & 124.6091 \\
allene             &  0.0008 &  0.0001&  0.0009 & 0.2681 &  0.0009 &  0.2526 & 1.3231 & 120.8339 \\
ammonia            &  0.0010 &  0.8635&  0.0010 & 0.8699 &  0.0010 &  0.8638 & 1.0235 & 103.9052 \\
benzene            &  0.0022 & -0.0001&  0.0480 & 0.5909 &  0.0004 &  0.0020 & 1.4061 & 120.0001 \\
cyclopropane       &  0.0045 &  0.1133&  0.0063 & 0.3675 &  0.0042 &  0.3575 & 1.5144 & 117.7845 \\
Ethane (staggered) &  0.0064 &  0.0020&  0.0132 & 0.0756 &  0.0053 &  0.0038 & 1.5293 & 111.2640 \\
Ethane (eclipsed)  &  0.0067 &  0.0066&  0.0145 & 0.1831 &  0.0054 &  0.0088 & 1.5429 & 111.7995 \\
Ethylene (ethene)  &  0.0019 &  0.0000&  0.0026 & 0.3267 &  0.0026 &  0.3024 & 1.3455 & 121.3734 \\
formaldehyde       & -0.0072 &  0.0001& -0.0060 & 1.1569 & -0.0060 & -0.5770 & 1.2159 & 122.4430 \\
furan              &  0.0012 & -0.0434&  0.0439 & 0.8304 &  0.0012 &  0.0041 & 1.4363 & 133.4318 \\
methane            &  0.0067 & -0.0000&  0.0067 & 0.0068 &  0.0067 &  0.0003 & 1.0995 & 109.4712 \\
propane            &  0.0069 & -0.0035&  0.0138 & 0.8381 &  0.0052 & -0.1038 & 1.5304 & 112.3029 \\
pyrrole            &  0.0035 & -0.6714&  0.0475 & 1.9507 &  0.0019 & -0.6106 & 1.4270 & 131.2667 \\
water              & -0.0026 &  1.6053& -0.0026 & 1.6053 & -0.0026 &  1.6053 & 0.9647 & 101.9080 \\
\midrule
ME                 & 0.0024 & 0.1339& 0.0144 & 0.734 & 0.0022 & 0.1472 & & \\
MAE                & 0.0038 & 0.2365& 0.0157 & 0.734 & 0.0034 & 0.3386 & & \\
SD                 & 0.0039 & 0.5216& 0.0185 & 0.5873 & 0.0034 & 0.5562 & & \\
SD-pop             & 0.0038 & 0.5026& 0.0178 & 0.5659 & 0.0033 & 0.5360 & & \\
\midrule
RMSE               & 0.0045 & 0.5202&  0.0229 & 0.9269 & 0.0040 & 0.5558 & & \\
MPE (\%)           & 4.3073 & 1.6490& 15.2243 & 7.2509 & 4.0368 & 2.8217 & & \\
\bottomrule
\end{tabular}
  \begin{tablenotes}
    \item[1]
$\mathrm{ME} = \sum_i^N \frac{\Delta x_i}{N}$,
$\mathrm{MAE} = \sum_i^N \frac{|\Delta x_i|}{N}$,
$\mathrm{RMSE} = \sqrt{\sum_i^N \frac{(\Delta x_i)^2}{N}}$,
$\mathrm{MPE} = \frac{1}{N}\sum_i^N \frac{|\Delta x_i|}{|x_i^{\rm MRCC}|}\times 100$,
$\mathrm{SD} = \sqrt{\frac{\sum_i^N \left(\Delta x_i - \overline{\Delta x}\right)^2}{N-1}}$,
$\mathrm{SD\mbox{-}pop} = \sqrt{\frac{\sum_i^N \left(\Delta x_i - \overline{\Delta x}\right)^2}{N}}$,
where $x$ denotes either a bond length $r$ or a bond angle $\theta$ and $\overline{\Delta x}$ is the mean error over the data set.
  \end{tablenotes}
\end{table*}

\begin{table}[t]
\caption{Root-mean-square errors (RMSE) of OOpCCD structural parameters at equilibrium geometries relative to MP2 reference structures using a cc-pVDZ basis set.
    The mean RMSE without rings is calculated for the same set of molecules, excluding molecules with aromatic rings, that is, benzene, furan, and pyrrole.
    $R$: Bond-length errors in \AA.
    $\angle$: bond-angle errors in degrees.
    A similar analysis is performed for CCSD(F12c)(T*) reference data, which is summarized in the SI.
    }
\label{tab:1_rmse-per-molecule}
\centering
\renewcommand{\arraystretch}{1.15}
\setlength{\tabcolsep}{6pt}
\begin{tabular}{l rr}
\toprule
    Molecule & \multicolumn{2}{c}{RMSE} \\
    & R [\AA] & $\angle$ [$^\circ$]
    \\
\midrule
Acetaldehyde        & 0.0079 & 0.5764 \\
Allene              & 0.0008 & 0.2612 \\
Ammonia             & 0.0010 & 0.8635 \\
Benzene             & 0.0312 & 0.4785 \\
Cyclopropane        & 0.0048 & 0.5064 \\
Ethane (staggered)  & 0.0070 & 0.0672 \\
Ethane (eclipsed)   & 0.0074 & 0.1751 \\
Ethylene            & 0.0023 & 0.4409 \\
Formaldehyde        & 0.0073 & 0.8180 \\
Furan               & 0.0197 & 0.6054 \\
Methane             & 0.0067 & 0.0053 \\
Propane             & 0.0077 & 0.3175 \\
Pyrrole             & 0.0242 & 1.7717 \\
Water               & 0.0026 & 1.6053 \\
\midrule
Average ME          & 0.0032 & -0.0323 \\
Average MAE         & 0.0094 &  0.4524 \\
Average RMSE        & 0.0158 &  0.7143 \\
Mean RMSE w/o rings & 0.0061 &  0.4266 \\
\bottomrule
\end{tabular}
\end{table}

\begin{figure*}[!t]
\centering
\includegraphics[width=\linewidth]{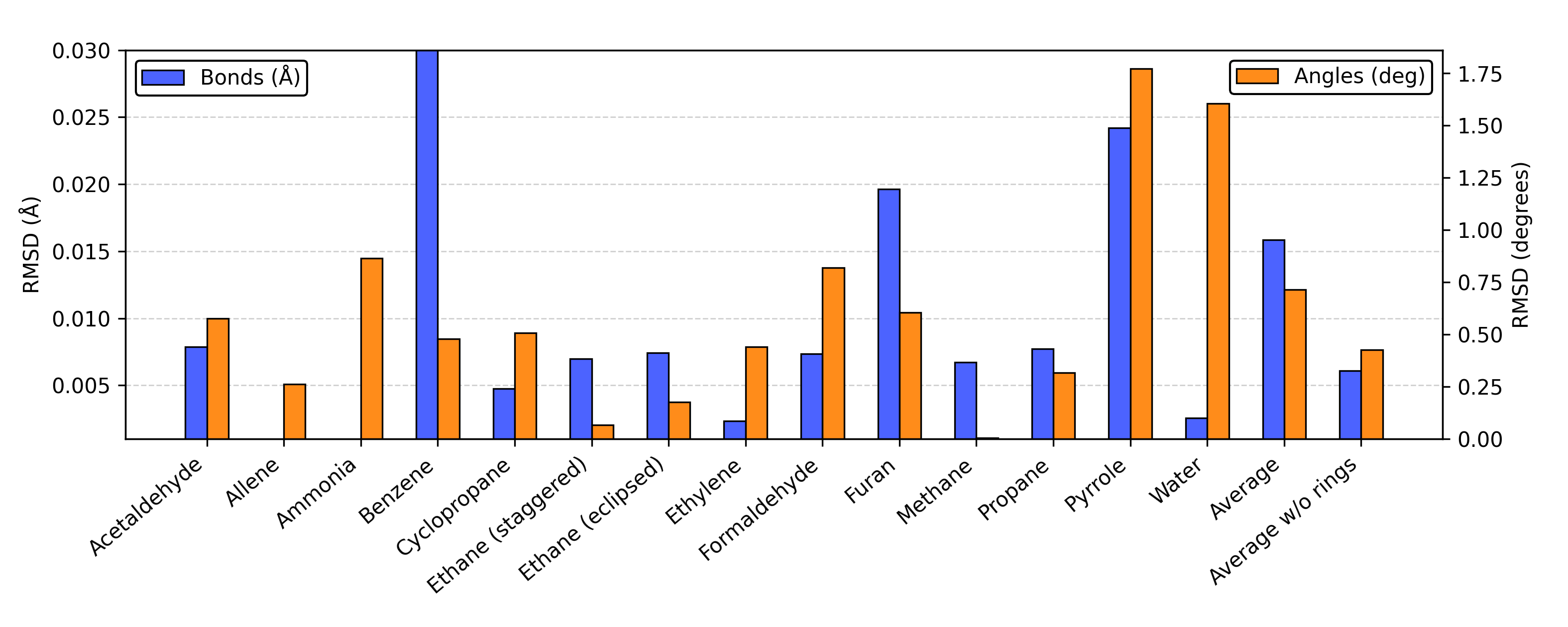}
\caption{
RMSEs of bond lengths (\AA, blue bars) and bond angles (degrees, orange bars) for OOpCCD equilibrium geometries relative to CCSD(F12c)(T*) reference structures, with all OOpCCD calculations performed using the cc-pVDZ basis set.
The right vertical axis refers to angular deviations (degrees), while the left vertical axis corresponds to bond-length deviations (\AA).
The mean RMSE values (excluding aromatic systems) are 0.0061~\AA\ for bond lengths and 0.4266$^\circ$ for bond angles.
}
\label{fig:rmse-two-axis}
\end{figure*}

\subsection{Molecular structure benchmark calculations}
\label{sec:geom-errors}

To quantify the accuracy of analytic OOpCCD geometry optimizations, we compare selected equilibrium structural parameters against MP2 and CCSD(F12c)(T*) reference geometries taken from Ref.~\citenum{crisci2025energies}.
For each molecule in the test set, we report the maximum, average, and median deviations in bond lengths and bond angles, determined for all structural parameters $i$ for each molecule $M$ as
\begin{align}
\Delta r_{M_i} &= r_{M_i}^{\rm OOpCCD} - r_{M_i}^{\rm reference}, \\
\Delta \theta_{M_i} &= \theta_{M_i}^{\rm OOpCCD} - \theta_{M_i}^{\rm reference}, 
\end{align}
where $r$ is given in \AA\ and $\theta$ in degrees.
In addition to the per-molecule maxima, averages, and medians, we summarize the overall performance using the following statistical metrics: mean error (ME), mean absolute error (MAE), root-mean-square error (RMSE), mean percentage error (MPE), and standard deviation (SD).
For completeness, we also provide the population standard deviation (SD-pop).
These measures are collected and defined in Table~\ref{tab:geom-errors} for each (maximum, average, and median) deviation in bond lengths and angles for the MP2 reference structures.
The RMSE values for all bond lengths and angles are collected in Table~\ref{tab:1_rmse-per-molecule}.
A graphical illustration of the RMSE is shown in Figure~\ref{fig:rmse-two-axis}.

In general, OOpCCD predicts structural parameters that are in good agreement with MP2 reference data, despite OOpCCD missing a significant fraction of dynamical correlation.
The average ME for bond lengths is around 0.003 \AA{} and increases to 0.009 \AA{} when taking the absolute counterpart (MAE).
The large average RMSE value of about 0.016 \AA{} is due to the observed symmetry-breaking in all aromatic systems.
It is a known problem that electron-pair methods, including OOpCCD, destroy aromaticity and lead to uneven C--C bond lengths.\cite{boguslawski2014nonvariational,tailored-pccd-jctc-2022}
Excluding aromatic rings from the test set reduces the RMSE to 0.006 \AA{}.

The deviations in bond angles feature a larger spread than observed for bond lengths.
The average ME for the complete test set is small and amounts to only 0.03$^\circ$, increasing to 0.45$^\circ$ for the absolute counterpart.
The RMSE further increases to 0.71$^\circ$, which can be attributed to larger bond angle deviations in molecules featuring lone pairs, like ammonia, formaldehyde, pyrrole, or water.
Most likely, OOpCCD predicts more localized lone pairs, which results in larger interatomic bond angles.
On the other hand, OOpCCD both over- and underestimates bond lengths in molecules featuring lone pairs (cf., ammonia for overestimation and water for underestimation).
The error measures for the median (see Table~\ref{tab:geom-errors}) point to a rather small spread in the structural parameter deviations (around 0.003 \AA{} or 0.3$^\circ$), suggesting that OOpCCD can predict robust structural parameters.
On the contrary, the maximum error measures increase to 0.02 \AA{} and up to 1$^\circ$ due to known aromatic ``outliers''.
Finally, we should note that OOpCCD breaks linearity as we do not impose any symmetry constraints during the optimization (to ensure localization of all orbitals).
The deviation from linearity is, however, small and amounts to less than 0.2$^\circ$ (allene), which is still below the statistical average.

Similar observations can be made when comparing the OOpCCD-predicted structural parameters (bond lengths and bond angles) to CCSD(F12c)(T*) reference data (determined using numerical gradients and summarized in the SI).
In contrast to MP2 reference data, the ME and MAE increase to approximately 0.014 and 0.017 \AA{}, respectively, pointing to an elongation of interatomic distances in OOpCCD due to the lack of dynamical correlation.
However, the deviations are more consistent across the test set, where removing the aromatic cases from the error statistics only marginally affects the error measures.
The errors in bond angles, on the other hand, are rather similar to the MP2 analysis above.
The MAE slightly increases to 0.5$^\circ$, while the RMSE amounts to 0.76$^\circ$.
The major difference is that bond angles are underestimated within OOpCCD when the system features lone pairs (like water and ammonia), while they are overestimated with respect to MP2 reference data.

\begin{figure*}[!t]
\centering
\includegraphics[width=0.95\textwidth]{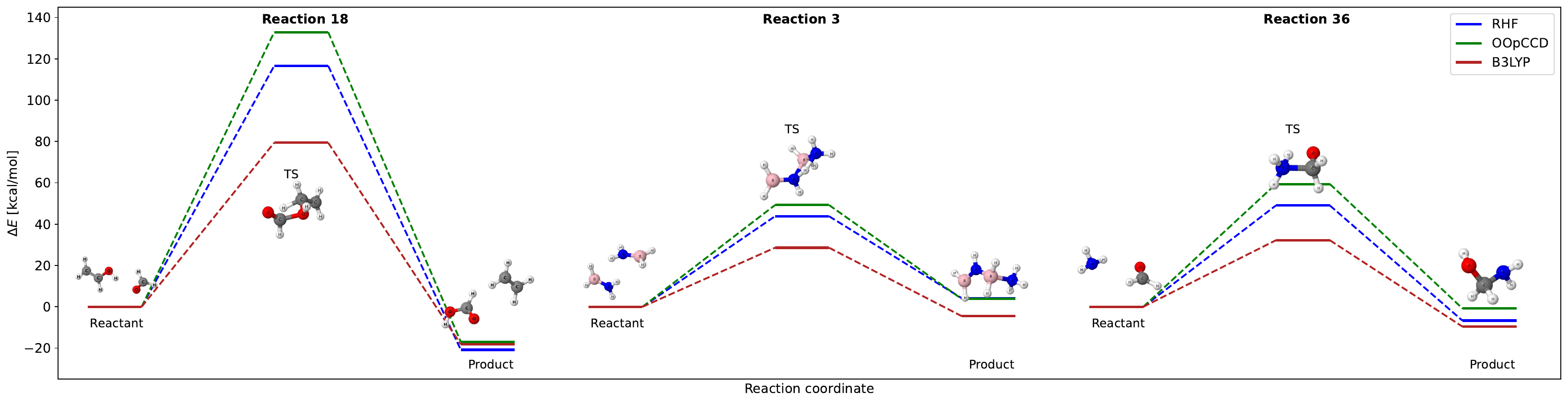}
\caption{Simplified reaction coordinate of \emph{reaction 18}, {\emph{reaction 3}, and \emph{reaction 36}} discussed in Ref.~\citenum{asgeirsson2021nudged}.
The structures shown are the OOpCCD-optimized ones.
All molecular structures are visualized using the Jmol package.\cite{jmol}}
\label{fig:pes-ts-search}
\end{figure*}

\subsection{Transition-state optimization tests}
\label{sec:ts-validation}

In addition to equilibrium-geometry optimizations, we assessed the performance of OOpCCD for transition-state (TS) searches.
Transition-state optimization is generally more challenging than minimum-energy optimization because the algorithm must converge to a first-order saddle point on the potential-energy surface rather than to a local minimum.\cite{schlegel2011geometry,hratchian2005finding}
As a result, the convergence behavior can be sensitive to the initial guess geometry, the coordinate representation, and the step-control strategy used by the optimizer.\cite{wang2016geometry}
{To tackle this task,} we chose the \emph{reaction 18} reaction example reported in Ref.~\citenum{asgeirsson2021nudged} as an initial test for our transition-state optimization algorithm.

Starting from a transition-state guess geometry, the RHF/cc-pVDZ optimization smoothly converged to a stationary point characterized by a single imaginary vibrational frequency, confirming that the structure corresponds to a first-order saddle point on the potential-energy surface.\cite{eriksen2021incremental}
On the contrary, TS optimization within OOpCCD proved more challenging as locating the TS strongly depends on the guess structure taken.
The OOpCCD-predicted TS could, however, be easily targeted starting from the RHF-preoptimized TS structure.

{To further assess the robustness and generality of the OOpCCD-based TS optimization, we extended the analysis to two additional reaction systems, denoted as \textit{reaction~3} and \textit{reaction~36} reported in Ref.\mbox{~\citenum{asgeirsson2021nudged}}.}
{For each system, full reaction profiles including reactant, transition-state, and product structures were determined and are summarized in Figure}~\ref{fig:pes-ts-search}{.}
{Across all three systems, OOpCCD correctly identifies first-order saddle points and reproduces the qualitative shape of the reaction profiles.}
{This demonstrates that the successful OOpCCD TS search for \textit{reaction~18} is not system-specific, but can be transferred to chemically distinct transition-state examples.}

Table~\ref{tab:ts-rhf-pccd} summarizes the RHF, OOpCCD, and B3LYP energies for the reactant, TS, and product structures {of all three reactions}.
For each method, the absolute imaginary frequency at the saddle point is also printed.
{The reaction energies, $\Delta E$(Product), predicted by OOpCCD for \textit{reaction~18} are in reasonable agreement with the B3LYP reference values.}
{For \textit{reaction~3} and \textit{reaction~36}, OOpCCD yields the correct qualitative ordering of stationary points but overestimates the product energies relative to B3LYP.}
{The absolute imaginary frequencies are also reproduced qualitatively, although OOpCCD systematically overestimates their magnitudes compared to B3LYP.}
{A more pronounced discrepancy is observed for the barrier heights, $\Delta E$(TS), which are consistently overestimated by OOpCCD for all three systems.}
{For example, in \textit{reaction~18}, the OOpCCD barrier height amounts to 132.9~kcal/mol compared to the B3LYP reference value of 79.4~kcal/mol, while similar overestimations are observed for reaction and \textit{reaction~36}.}
{This systematic trend indicates that dynamical correlation effects, which are not fully captured within the seniority-zero OOpCCD framework, play an essential role in accurately describing transition-state energetics and} barrier heights.
These effects can be straightforwardly included \textit{a posteriori}, exploiting the OOpCCD-optimized molecular structures.

{Taken together, these transition-state calculations demonstrate that OOpCCD analytic-gradient optimizations can robustly locate first-order saddle points, while quantitative barrier heights require additional dynamical-correlation corrections.}

\begin{table*}[!t]
\caption{Total $E_{\rm tot}$ [$E_h$] and relative $\Delta E$ [kcal/mol] energies for the structures involved in the transition-state test.
    All relative energies are given wrt the reactants.
    The absolute imaginary frequency at the saddle point $|\omega_0|$ [cm$^{-1}$] is also provided.
    All calculations were performed with a cc-pVDZ basis set.
    The B3LYP\cite{lyp1,becke88,b3lyp_becke,b3lyp} data was generated for the def2-SVP\cite{def2-svp,def2-svp-2} basis set and is taken from Ref.~\citenum{asgeirsson2021nudged}.
}

\label{tab:ts-rhf-pccd}
\centering
\setlength{\tabcolsep}{8pt}
\begin{tabular}{ll rrr rrr rr}
\toprule
 &  & \multicolumn{3}{c}{RHF}
    & \multicolumn{3}{c}{{OOpCCD}}
    & \multicolumn{2}{c}{B3LYP\cite{asgeirsson2021nudged}}\\
\cmidrule(lr){3-5}
\cmidrule(lr){6-8}
\cmidrule(lr){9-10}
Rxn & Structure
    & $E_{\rm tot}$ & $\Delta E$ & $|\omega_0|$
    & $E_{\rm tot}$ & $\Delta E$ & $|\omega_0|$
    & $\Delta E$ & $|\omega_0|$ \\

\midrule

   & Reactant & -163.001591 & 0.0 & -- & -163.284517 & 0.0 & -- & 0.0 & -- \\
3  & TS       & -162.931771 & 43.8 & 1866.3 & -163.206031 & 49.3 & 2354.1 & 28.6 & 1246.3 \\
   & Product  & -162.994898 & 4.2 & -- & -163.278418 & 3.8 & -- & -4.5 & -- \\
\\
   & Reactant    & -266.792920 & 0.0 & --     & -267.134450 & 0.0 & --     & 0.0 & -- \\
18 & TS          & -266.607075 & 116.6 & 1711.3 & -266.922614 & 132.9 & 1741.4 & 79.4 & 1081.4 \\
   & Product     & -266.826034 & -20.8 & --    & -267.161559 & -17.0 & --     & -18.2 & -- \\
\\
   & Reactant & -170.079368 & 0.0 & -- & -170.313857 & 0.0 & -- & 0.0 & -- \\
36 & TS      & -170.001250 & 49.0 & 1827.4 & -170.219202 & 59.4 & 1983.8 & 32.2 & 1507.7 \\
   & Product & -170.090021 & -6.7 & -- & -170.315134 & -0.8 & -- & -9.6 & -- \\

\bottomrule
\end{tabular}
\end{table*}



\section{Conclusions}\label{sec:conclusion}

In this work, we derived and implemented a reusable geometry-optimization engine within the \textsc{PyBEST} software package~\cite{pybest-paper-2021,boguslawski2024pybest} and interfaced it with the \texttt{geomeTRIC} optimizer, thereby enabling analytic-gradient-driven molecular structure optimizations within a robust TRIC-based framework.
Specifically, we implemented the analytic nuclear gradients for orbital-optimized pair coupled-cluster doubles (OOpCCD/AP1roG).
The gradient expressions are derived within a Lagrangian formalism and formulated in terms of response density matrices and generalized Fock matrix.
Provided no frozen-core approximation is employed, explicit orbital-response terms are avoided, resulting in a compact implementation.

The performance of the OOpCCD \textsc{PyBEST}--\texttt{geomeTRIC} workflow was assessed for equilibrium geometries and transition states.
For diatomic molecules, the optimized bond lengths and energies obtained from analytic gradients are in very good agreement with reference values extracted from fitted potential-energy curves, confirming the correctness of the gradient implementation (in addition to comparisons to numerical gradients).
For the set of small closed-shell molecules, {OOpCCD} yields equilibrium structures that are generally in good agreement with MP2 and CCSD(F12c)(T*) reference geometries, despite the lack of dynamical correlation within the OOpCCD model.
While bond-length deviations remain small for most systems, larger discrepancies are observed for aromatic molecules, reflecting the known limitations of electron-pair methods in describing (aromatic) delocalized $\pi$-systems.
Specifically, OOpCCD yields (typically overestimated) bond lengths with an RMSE of 0.02 \AA{} compared to CCSD(F12c)(T*) reference data, which is an acceptable error considering the lack of dynamical correlation in the pCCD model.
Bond-angle deviations are typically below 1$^\circ$, with the largest deviations observed for systems containing lone pairs.

We further demonstrated the applicability of the optimization engine to transition-state searches.
The RHF-based workflow robustly converges to a first-order saddle point, whereas OOpCCD calculations are more sensitive to the initial guess geometry in {the} studied {cases}.
Nevertheless, starting from an RHF-preoptimized structure enables the successful location of the corresponding OOpCCD transition {states}.
The resulting reaction energies and imaginary frequencies are in reasonable agreement with reference data, although the barrier height is overestimated, indicating the importance of dynamic correlation for quantitative accuracy.

Overall, the present work establishes a robust and modular optimization framework in \textsc{PyBEST} for analytic molecular gradients and geometry optimizations with seniority-zero wavefunctions.
The developed infrastructure provides a general platform for future extensions toward more advanced correlated methods and applications, including improved equilibrium structures, reaction-path calculations, and transition-state studies within the \textsc{PyBEST} environment.
Finally, our current implementations rely on dense representations of the electron repulsion integrals, which limit the applicability of OOpCCD geometry optimizations to small or medium-sized problems.
To tackle larger systems, these dense electron repulsion integrals and their derivatives need to be replaced by, for instance, Cholesky-decomposed counterparts.
Analytic gradient evaluation for a Cholesky-driven OOpCCD framework and its efficient implementation will be focus of future work.

\section{Supplementary Material}
The \texttt{.zip} file contains the (i) \texttt{SI\_mp2.xlsx} file, which collects the complete list of the OOpCCD-optimized structural parameters and differences wrt MP2 data, including the statistical error analysis, (ii) \texttt{SI\_cc.xlsx} file, which collects the complete list of the OOpCCD-optimized structural parameters and differences wrt CCSD(F12c)(T*) data, including the statistical error analysis, (iii) \texttt{SI.pdf} file, which visualizes the error measures wrt CCSD(F12c)(T*) data {and contains selected OOpCCD wall times}.
All OOpCCD-optimized molecular structures are attached as \texttt{.xyz} files.

\section{Data Availability Statement}
The data that supports the findings of this study are available within the article and its supplementary material.

\section{Conflicts of Interest}
There are no conflicts to declare.

\begin{acknowledgments}
We gratefully acknowledge Polish high-performance computing infrastructure PLGrid (HPC Center: ACK Cyfronet AGH) for providing computer facilities and support within computational grant no.~PLG/2025/018840.
We thank Bastiaan van Hoorn for suggesting an interesting, but feasible reaction for the TS search included in this work. 
Funded/Co-funded by the European Union (ERC, DRESSED-pCCD, 101077420).
Views and opinions expressed are, however, those of the author(s) only and do not necessarily reflect those of the European Union or the European Research Council. Neither the European Union nor the granting authority can be held responsible for them.
\end{acknowledgments}

\bibliography{rsc}

\end{document}